\documentclass[]{spie}  

\usepackage{amsmath,amsfonts,amssymb}
\usepackage{graphicx}

\usepackage[colorlinks=true, allcolors=blue]{hyperref}

\title{Pandeia: A Multi-mission Exposure Time Calculator for JWST and WFIRST}

\author[a]{Klaus M. Pontoppidan}
\author[a]{Timothy E. Pickering}
\author[a]{Victoria G. Laidler}
\author[a]{Karoline Gilbert}
\author[a]{Christopher D. Sontag}
\author[a]{Christine Slocum}
\author[a]{Mark J. Sienkiewicz}
\author[a]{Christopher Hanley}
\author[a]{Nicholas M. Earl}
\author[a]{Laurent Pueyo}
\author[a]{Swara Ravindranath}
\author[a]{Diane M. Karakla}
\author[a]{Massimo Robberto}
\author[a]{Alberto Noriega-Crespo}
\author[a]{Elizabeth A. Barker}
\affil[a]{Space Telescope Science Institute, 3700 San Martin Dr., Baltimore, MD, USA}

\authorinfo{Further author information: (Send correspondence to T.E.P.)\\
T.E.P.: E-mail: pickering@stsci.edu, Telephone: 1 410 338 2460\\
K.M.P.: E-mail: pontoppi@stsci.edu, Telephone: 1 410 338 4744}

\begin{document}
\maketitle

\begin{abstract}
Pandeia is the exposure time calculator (ETC) system developed for the James Webb Space Telescope (JWST) that will be used for creating JWST proposals. It includes a simulation-hybrid Python engine that calculates the two-dimensional pixel-by-pixel signal and noise properties of the JWST instruments. This allows for appropriate handling of realistic point spread functions, MULTIACCUM detector readouts, correlated detector readnoise, and multiple photometric and spectral extraction strategies. Pandeia includes support for all the JWST observing modes, including imaging, slitted/slitless spectroscopy, integral field spectroscopy, and coronagraphy. Its highly modular, data-driven design makes it easily adaptable to other observatories. An implementation for use with WFIRST is also available.
\end{abstract}


\section{Introduction} \label{sec:intro}

Exposure time calculators (ETCs) are integral parts of any large observatory. They are often one of the first places new users go to explore and familiarize themselves with the capabilities of a new facility and one of the last tools that are used before submitting an observing program for evaluation by a time allocation committee. As such, there is a strong drive to develop ETCs that accurately model a telescope/instrument combination as well as provide the user with an intuitive, easy-to-use, yet powerful interface. The JWST ETC system -- Pandeia -- is the environment in which a user will explore an observing concept before deciding to write a proposal and crafting specific observation sequences.

During the JWST era users will benefit from features that go well beyond what has previously been offered as part of an exposure time calculator. Such features include 1) algorithms that model the data acquisition, structure, and post-processing, 2) an advanced, interactive graphical user interface (GUI), 3) more powerful tools to visualize a large parameter volume, and 4) functionality to share work efficiently with collaborators.

The JWST ETC is required to support instrument modes that generate large amounts of information, including integral field unit (IFU) data and multi-object spectroscopy with the NIRSpec Multi-Shutter Array (MSA). Calculations must be accurate for modern observational modes that depend on high signal-to-noise and high contrast data, including exoplanet spectroscopy, coronagraphy, and non-redundant aperture masking.

A common property of JWST observations is that they are recorded as sets of two-dimensional detector exposures. All information is at some point stored in discrete detector pixels. Different observing modes fundamentally differ in two aspects: First, how light is projected and/or dispersed onto a detector and second, how observational products are extracted from a set of exposures, typically by a linear combination of pixel values. The ETC generalizes both of these steps by adopting a three-dimensional, pixel-based approach. The user can create complex astrophysical scenes consisting of multiple sources with different spectral and spatial properties. By modeling individual detector pixels the ETC can explicitly mimic many of the steps taken in the calibration and reduction of actual JWST data, leading to more accurate results that are easier to test and interpret. To facilitate extensive forward-model parameter explorations, the ETC engine is optimized for calculations of small ``postage-stamp'' regions on angular scales of a few arcseconds. This approach can generally lead to processing times as fast as a few seconds per calculation, depending on observing mode.

Key features of the Pandeia ETC include: 
\begin{itemize}
\item Full support of a wide variety of imaging, coronagraphic, and spectroscopic instrument modes.
\item Support for MULTIACCUM readout modes.
\item Includes effects of correlated noise and inter-pixel capacitance.
\item Advanced astronomical scene creation. 
\item Interfaces to the full JWST background model.
\item Fully data driven with verified reference data, including throughputs, point spread functions, and detector properties.
\item Built upon an extensible and modular Python engine. 
\end{itemize}
While originally developed for JWST, Pandeia has also been extended for use with the WFIRST wide-field imaging and integral-field spectroscopy instruments.

\section{Basic Philosophy and Concept}
\subsection{Features of a Pixel-based Approach}
There are a number of immediate advantages of a pixel-based ETC that cannot be easily modeled by a one-dimensional approach:
\begin{itemize}

\item The flux from a celestial source is not evenly divided among pixels, but follows a sharp point spread function. That is, the central pixel for a point source could be in a photon limited noise regime while the outer pixels are read noise dominated. This is further complicated by reduction methods that employ a weighting scheme, such as the commonly used optimal extraction.

\item The integration time for all pixels will not always be the same in MULTIACCUM integrations. Specifically, a bright source that reaches saturation during an integration may still produce sensible ramps in all pixels, as long as a minimum number of groups are unsaturated. In a long ramp with 100 groups, for example, the core of a point source could have an effective exposure time that is 1/50 that of the surrounding pixels. Accurately determining the total signal-to-noise therefore requires exact knowledge of the electron rate in each pixel separately.

\item In modern infrared detectors, the noise for each pixel is not independent. The biggest noise term for low flux NIR observations will be $1/f$ noise. The correlation follows a complex pattern and is stronger along the fast-read direction of the detector.

\item The pixel that detects the arrival of a photon is not always the pixel where the photon creates an electron-hole pair. The pixels do not have sharp boundaries as in a CCD and this smoothing affects many things when the image is under-sampled. So-called Inter Pixel Capacitance (IPC) causes an electronic coupling between adjacent pixels that affects the observed flux and noise in a non-Poissonian manner.

\item Observing strategy affects the signal-to-noise ratio of measured quantities. Selecting a detailed strategy for background subtraction will be an inherent, but highly variable, property of almost any JWST observation, and in many cases this will be done on a pixel-by-pixel basis.

\item Many observations seek to extract a source signal from a complex scene which may contain light from multiple sources. A pixel-based ETC will allow a user to investigate the sensitivity of planned observations to contaminating light from other sources in the field of view.

\end{itemize}

\subsection{Scope of Pandeia}

The Pandeia ETC distinguishes itself from a complete instrument simulator, although some parts of its functionality are similar. It inevitably represents a trade-off between realism, cost, and user experience. At the most basic level, the goal of Pandeia is to determine the signal-to-noise ratio of observables, as well as other ETC products, with sufficient accuracy to allow users to adequately plan and construct observations using the Astronomer’s Proposal Tool (APT). When adding complexity to the ETC model, prioritization is given to aspects that significantly affect the accuracy of its SNR prediction. Performance (i.e., the speed of a single computation) is also a significant and real concern. If individual ETC calculations take too long to execute, users will struggle to properly assess their parameter space. Given the high data rate of MULTIACCUM observations with large-format detectors (NIRCam will produce almost 42 million pixels of information in a single exposure), Pandeia will not consider full instrument fields of view.  It will instead perform calculations over relatively small “postage-stamp” fields of view, typically a few arcseconds in size. Consequently, certain simulation-like aspects are generally considered out-of-scope for Pandeia:

\begin{itemize}

\item Pandeia assumes that the covariance noise sum holds (see section 8.2). Specifically, it is assumed that pixel arithmetic is always linear. One consequence of this is that Pandeia will not consider optical field distortion, at least not explicitly, and does not model the exact boundaries of instrument apertures. However, this information is typically accessible from APT. 

\item Pandeia does not include diffraction effects not handled implicitly by a pre-computed PSF library. Further, Pandeia does not consider PSF variations across a field of view, with the exception of coronagraphic modes.

\item Because field distortion is ignored, Pandeia will generally not know where individual, identifiable pixels fall relative to an astronomical source (bad pixels, for instance). It is assumed that dithers will mitigate bad pixels.

\item Pandeia is deterministic so that it always produces the same output, given identical input. Although there are stochastic aspects of a real observation, such as cosmic ray impacts, Pandeia will handle such effects in an average sense.

\end{itemize}

\section{Implementation}
\subsection{Astronomical Scenes}

A basic concept of Pandeia, and a requirement of a full three-dimensional treatment, is the construction of an astronomical scene. A scene is a specification of an idealized (before being observed by a telescope) spatial (two angular coordinates) and spectral brightness distribution. A scene is composed of a finite number of celestial sources. Each source may be a point source of infinitely small spatial extent, or it may be extended and follow a prescribed spatial brightness distribution. Each source is also associated with a spectral energy distribution. It is assumed that the spectral and two spatial dimensions are independent. That is, an extended source will have the same normalized spectrum along every line-of-sight. However, extended sources with more complex spatio-spectral structure may be realized by combining multiple sources. An example could be an extended galactic disk composed of an old stellar population adorned with compact or point-like star-forming regions.

A scene may be as simple as a single on-axis point source or as complex as a cluster of densely spaced high redshift galaxies, each with its own unique spectrum. The general default of an ETC calculation is that of a single on-axis point source. The complex versatility of scene creation will only need to be accessed by users interested in more detailed configurations.

Scenes are not intended to be models of the entire instrument field-of-view. They are defined on relatively small spatial “postage stamps”, typically squares of 3-20 arcseconds on the side. This restriction is not fundamental, but is put in place both to limit the computational resources needed and the time required for a single calculation. Users interested in the full instrument aperture should consider effects not modeled by the ETC, such as field distortion. The design of the user interface would also be affected if a much larger field of view were considered. If it is important to model the full instrument field of
view in one calculation, a full-fledged instrument simulator is likely required. The sizes of the postage stamps are therefore chosen as a trade-off between computational time needed and typical use cases.

An astronomical scene is defined as a parameterized set of sources and associated spectra. This set of brightness parameters and ideal source spectra is called the abstract scene. Because the abstract scene is not dependent on the telescope or instrument, it can, once defined, be shared between calculations and observing modes. The abstract scene can be transformed into a realized scene. The realized scene cube is, at the time of realization, sampled on the pixel grid relevant to a given observing mode in order to avoid computationally intensive interpolations later on in the ETC calculation.

\begin{figure}[ht!]
\centering
\includegraphics[width=8cm]{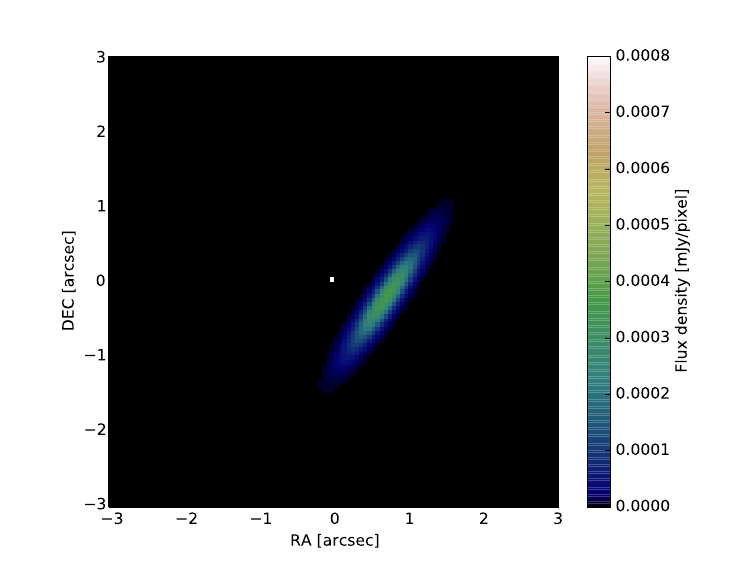}
\includegraphics[width=8cm]{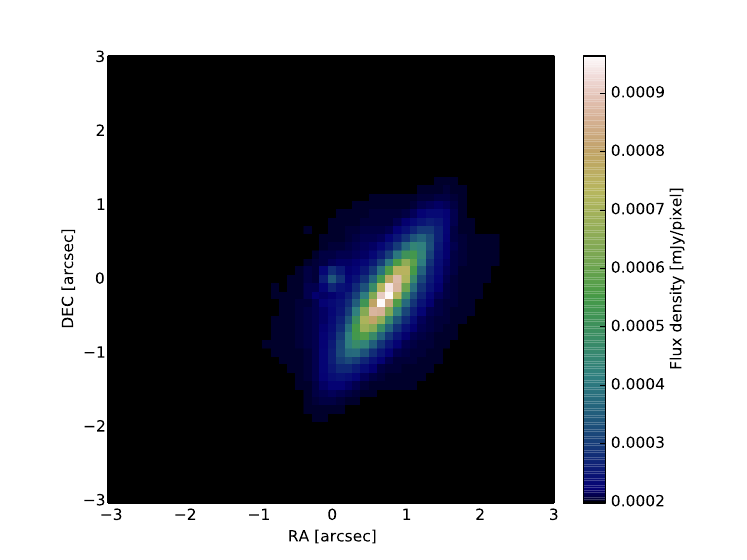}
\caption{Projection of an abstract scene cube for imaging ({\em left}) and the corresponding realized cube ({\em right}).}
\label{Scene}
\end{figure}

Realized scenes are initially spatially sub-sampled with an integer factor relevant to the detector plate scale. This is done as a trade-off to avoid undersampling and aliasing, yet allow fast and efficient resampling to a detector scale without expensive and potentially inaccurate interpolations.  An example of a monochromatic abstract and realized scene plane is shown in Figure \ref{Scene}.

\subsection{The Scene Coordinate System}

The abstract scene is specified in coordinates relative to the geometric scene center, $(\Delta X,\Delta Y)$. The realized scene is notionally rotated to the aperture angle of the relevant instrument mode. The pixel grid is assumed aligned with the aperture
angle, which is generally a good approximation. The realized scene coordinates are called $(\xi,\eta)$. For spectroscopic modes, $\xi$ is aligned with the dispersion axis, while $\eta$ is aligned with the cross-dispersion axis. In general, the scene rotation corresponds to an angle on the sky, but the ETC itself does not provide a measure of this angle, or the angle relative to the telescope coordinate system.

Any observation of a scene is a representation of some aspect of a realized scene cube. The realized scene cube is a brightness distribution, with a given sampling $(i,j,k)$:

\begin{equation}
I_{\rm scene} = I_υ (\xi_i,\eta_j,\lambda_k )  [\mathrm{mJy/pixel}]
\end{equation}

\subsection{Source Creation}

Astronomical sources that define scenes are generally created by specifying two sets of parameters; spatial and spectral. The spatial parameters include extent (point source, Gaussian ellipsoidal, Sers\'ic profile, or spatially flat) and spatial offset from the center of the postage stamp field of view of the scene. The location of the scene field of view relative to the instrument aperture is not explicitly modeled, but is implicitly assumed via the PSF and the configuration used to create it. In general, it is reasonable to assume the scene is centered in the instrument aperture. Whether there is a need for a more explicit model of the optical location of the scene is currently under review.  

A wide variety of source spectra are available for use including power-law, blackbody, stars (via a grid of Phoenix models\cite{1999ApJ...512..377H}), and integrated spectra of a range extragalactic objects\cite{0067-0049-212-2-18}. It is also straightforward to input a user-supplied spectrum. All spectra can be renormalized to given flux density at a specified reference wavelength or magnitude in a standard photometric system, including the JWST filters.

\subsection{Spectral Lines and other Spectral Modifications}

The source creation step includes functionality to add spectral lines. Lines are created by specifying a central (rest-frame) wavelength, intrinsic width parameters (in velocity space and in opacity for absorption lines), and an integrated line strength for emission lines or a central optical depth for absorption lines. Line profiles are currently modeled as Gaussian with implementation planned to generalize them to Voigt profiles (which are Gaussian in the limit of zero damping). Multiple lines can be added, either to model a complex line spectrum or to model complex line shapes (P Cygni line profiles, for instance). From the user perspective, line profiles are idealized to infinite resolution and any broadening due to the finite resolving power of the active observing mode is taken into account by the ETC without user input. This means that the sensitivity to under-resolved lines in the presence of strong continuum emission can be accurately modeled.

The source spectra will be further modifiable by velocity shifting a spectrum using an intrinsic Doppler shift and/or a cosmological redshift. Extinction may be added using one of a set of standard infrared extinction laws. \cite{2001ApJ...548..296W} \cite{2007ApJ...663.1069F} \cite{2009ApJ...691..560C}

\subsection{The PSF Library}

An integral part of Pandeia is a set of simulated point spread functions (PSFs) organized in a pre-computed PSF library (see Figure \ref{Scene}). The PSF library consists of monochromatic PSFs calculated using WebbPSF. \cite{2012SPIE.8442E..3DP}\cite{2014SPIE.9143E..3XP} Separate sets of PSFs are included for each instrument mode, relevant for the center of each instrument aperture. Thus, in most cases, the PSF profile is assumed to be constant across the field of view, with the exception of coronagraphic modes (see section 12.0). The PSFs are sampled on the same grid as the realized scene cubes, which in turn are subsampled relative to the detector pixel scale by an integer factor. The library PSFs are normalized to one incident photon prior to any one-dimensional throughput corrections, but including losses due to diffractive optical elements, such as a pupil.

The primary use of the PSF library is as input kernels for convolution of the realized scene cube. Each plane is convolved with an appropriate monochromatic, but wavelength-dependent, PSF. Stopping short of a full optical model of the observatory, this step makes a number of simplifying assumptions. Since a pre-computed library cannot provide PSFs at every unique wavelength, a PSF for a given plane is typically linearly interpolated between adjacent library PSFs along the wavelength axis. This assumption limits the fidelity of the speckle structure of the assumed realization of the wave-front error. Another important assumption is that the PSF profile is independent of spatial location within the postage stamp field-of-view. Exceptions to this assumption are coronagraphic modes, where the occulted PSF is very different from the off-axis PSF, by design. For coronagraphic modes, each source is convolved with a subset of the PSF library relevant to the position of that source in the field-of-view. 

\section{Scene Cube Projections}

\begin{figure}[ht!]
\centering
\includegraphics[width=14cm]{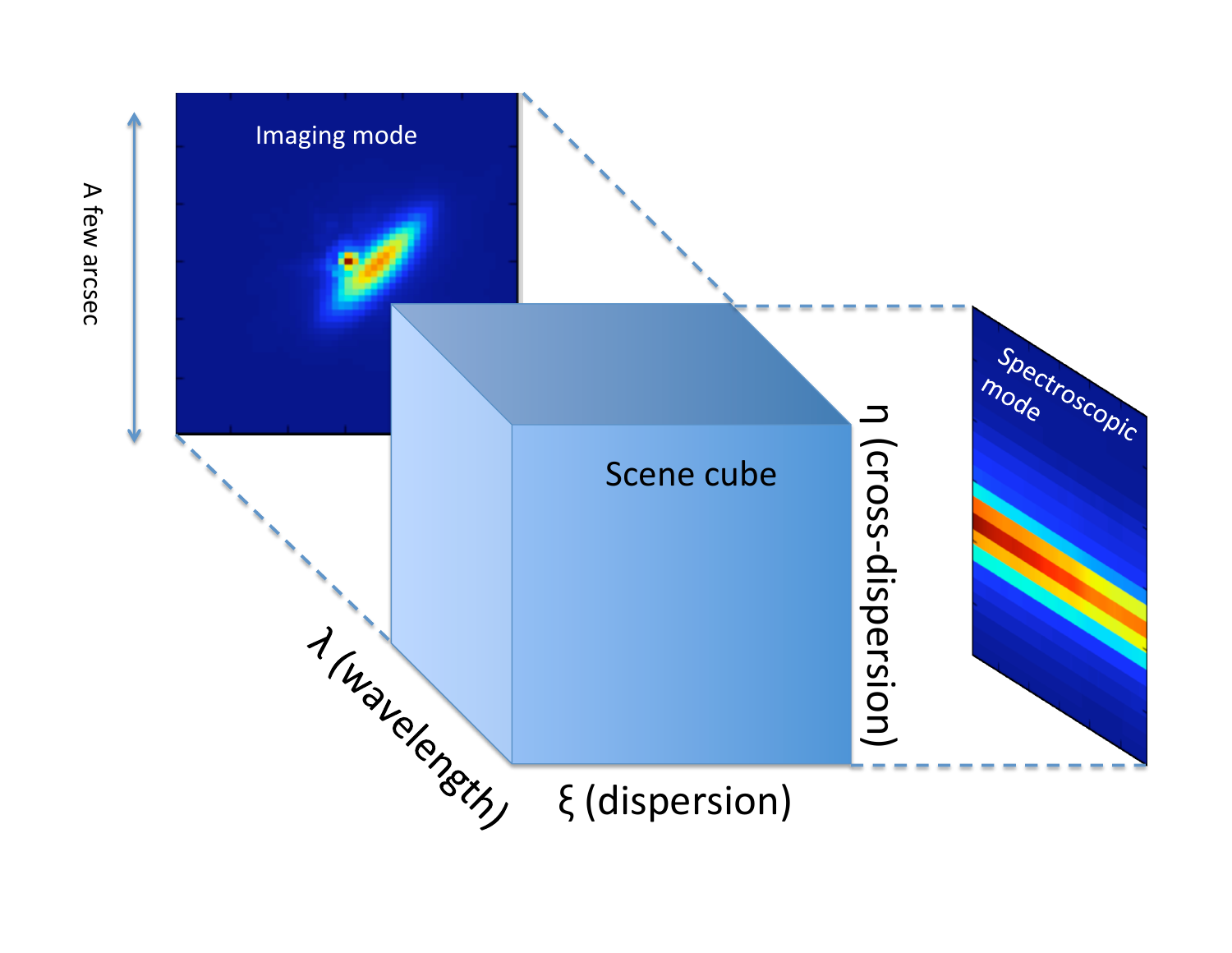}
\caption{}
\label{SceneCube}
\end{figure}

In order to calculate the electron rate per pixel on a detector, the realized scene cube must be projected onto a virtual detector plane according to the active observing mode. The projection is generally a map from scene cube space to detector space:

\begin{equation}
P: S(\xi, \eta, \lambda)\rightarrow F(x,y)
\end{equation}

where $(x,y)$ is the detector pixel coordinate system and $F(x,y)$ is a rate in units of electrons/pixel. Consider, for instance, an imaging mode: In this case, the cube is integrated along the wavelength axis to create a two-dimensional image. Conversely, a spectroscopic mode is projected onto the detector by integrating along the spatial axis corresponding to the dispersion or cross-slit, or cross-slice (in the case of an IFU) direction. This basic concept is illustrated in Figure \ref{SceneCube}. There are currently five different types of scene projection in the ETC: Imaging, single slit/slitlet, IFU, slitless, and multi-order
slitless.

\subsection{Imaging Projections}
An imaging-mode detector plane is calculated by integrating the scene rate along the wavelength axis:

\begin{equation}
F(x,y) = \int{S_{\lambda}^{\rm rate} (\xi ,\eta ,\lambda)d\lambda} \quad {\rm [electrons/s/pixel]}
\end{equation}

The numerical integration uses a quadratic Newton-Cotes formula (Simpson's Rule). While the calculation of realized scene cubes is somewhat different for coronagraphic imaging, the projection is the same. NIRISS non-redundant aperture mask images are also rendered using the standard imaging mode projection.

\subsection{Fast Projection for Narrow-Slit Spectroscopy}

Single-slit spectroscopy (NIRSpec MSA, fixed slit, and MIRI LRS) is implemented by applying a geometric slit mask, $M_{\rm slit}(\xi,\eta)$, to the PSF-convolved scene. The masked scene can be integrated along the scene dispersion axis to produce a two-dimensional spectrum (one spatial and one wavelength axis):

\begin{align*}
F(\lambda,y) = \frac{d\lambda}{dx} \int S_{\lambdaλ}^{\rm rate} \times M_{\rm slit} (\xi,\eta)d\xi & \\
= \frac{d\lambda}{dx} \Sigma_i S_{\lambda}^{\rm rate} \times M_{\rm slit} (\xi_i,\eta_j) & \quad {\rm [electrons/s/pixel]}
\end{align*}

where $d\lambda$/$dx$ is the spectral dispersion function mapped on the detector plane in units of $\rm \mu m/pixel$. Because the wavelength grid is not sampled in pixel space at this point, the final step is to interpolate the detector rate on the correct wavelength/pixel scale using the dispersion function: $F(\lambda,y)\rightarrow F(x(\lambda),y)$. Note that because the PSF is already appropriately sampled on sub-pixel scales and the rebinning to the pixel scale is flux conserving, the integration can be replaced by a simple sum, by design.

The advantage of this approach is speed, while the disadvantage is that small spatial offsets between sources in the slit are not translated into spectral offsets on the detector. That is, the assumption is that the slit width is similar to or smaller than the PSF width. For very large slits or slitless spectroscopy, a different projection algorithm is needed.

\subsection{Fast Projection for Integral Field Spectroscopy}

The integral field units (IFUs) on JWST (NIRSpec IFU and MIRI MRS) and WFIRST use image slicers to create an array of spectra across the field of view. Pandeia treats each slice individually as a narrow slit and uses the fast narrow-slit projection to map the spectrum for each slice's section of the field of view onto a section of the detector plane. The resulting two-dimensional slices are stacked onto a single detector plane, which approximates what an IFU observation looks like in actual detector data, only without field distortion.

\subsection{Projection for Slitless or Wide-Slit Spectroscopy}

In the case of slitless spectroscopy or spectroscopic modes with a slit wider than the PSF core, the scene cube projection becomes computationally more intensive. Because each detector pixel typically receives light from all wavelengths within the spectroscopy band pass, the projection line integral is carried out over an oblique curve, $l$, through the scene cube:

\begin{equation}
F(x,y) = \int{S_{\lambda}^{\rm rate} (\xi,\eta,\lambda) dl}
\end{equation}

Each point in the scene cube is mapped uniquely to the detector. If we assume that the dispersion function $d\xi$/$d\lambda$ does not vary across the field of view, the curve can be parameterized as:

\begin{align}
\xi = \Delta \xi(\lambda)+\xi_0 (x,y) = \Delta \xi(t)+\xi_0 (x,y) \\
\eta = \eta_0 (x,y) \\
\lambda = t
\end{align}

and the line integral can be rewritten as:

\begin{equation}
F(x,y) = \int{S_{\lambda}^{\rm rate} (\xi(x,y,\lambda),\eta(x,y),\lambda) \sqrt{\Big(\frac{d\xi}{d\lambda}\Big)^2+1}  d\lambda}
\end{equation}

Note that in the limit of no dispersion, $d\xi/d\lambda=0$, the expression reduces to that of the imaging projection.

By definition, the slitless projection cannot include sources located outside of the scene postage stamp field of view. Background contributions from areas outside of the scene can be added appropriately because it can be assumed to not vary spatially. This means that the baseline concept for slitless spectroscopy will be able to handle single objects well. This is anticipated to be
the most common use of slitless spectroscopy for MIRI, NIRISS, and NIRSPEC (the large fixed slit aperture is considered a slitless
mode). The size of the scene in the dispersion direction for slitless modes is not fundamentally limited, but will be determined as a trade-off between science needs and performance.

\subsection{Projection for NIRISS Single-Object Slitless Mode}

\begin{figure}[ht!]
\centering
\includegraphics[width=16cm, height=2cm]{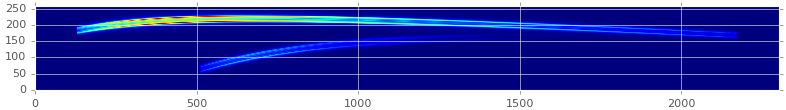}
\caption{Example output of a Pandeia calculation for NIRISS SOSS mode. This calculation includes 1st and 2nd order.}
\label{soss}
\end{figure}

The NIRISS\cite{2012SPIE.8442E..2RD} instrument on JWST includes a cross-dispersed single-object slitless spectroscopy (SOSS) mode that is optimized for single sources that are relatively bright. In this case, the scene cube is projected onto the detector multiple times, once for each of the supported orders. All orders must be included to properly account for overlapping background and contaminating signal. In addition, there is significant spatial distortion (see Figure \ref{soss}) that is unique to each order. The distortion is implemented as a "trace" function, $\mathcal{T}(S, \lambda)$, that shifts the input scene along the cross-dispersion axis in the detector plane. Including this effect yields a line integral of:

\begin{equation}
F(x,y) = \int{\mathcal{T}(S_{\lambda}^{\rm rate} (\xi(x,y,\lambda),\eta(x,y),\lambda), \lambda) \sqrt{\Big(\frac{d\xi}{d\lambda}\Big)^2+1}  d\lambda}
\end{equation}

Support for distortion of this form is available in general for all slitless projections, but is only currently required for SOSS. Other supported dispersed modes thus far have measured distortion of less than a pixel.

\section{Propagation Through Telescope and Instruments}

A central step in an ETC calculation is to convert the incident flux cube (before the telescope) to a photon rate before the detector and then to an electron rate after detection. This step is handled by a throughput conversion applied to the scene cube after PSF convolution.

\subsection{Instrument Configuration}

The ETC is configured for use by a specific instrument and observing mode by selecting an instrument and subsequently one of its observing modes. For instance, the available NIRSpec observing modes are MSA, fixed slit, and IFU spectroscopy. Each observing mode is associated with a defined set of parameters such as choice of filter or dispersing element.

\subsection{Exposure Configuration}

In the simplest ETCs, there exists an analytic one-to-one relation between SNR and exposure time. This is no longer the case for MULTIACCUM infrared detectors where the SNR-time relation is neither analytical nor unique. Consequently, Pandeia always performs individual, or sets of, forward calculations. That is, a unique MULTIACCUM exposure specification is taken as input:

\begin{itemize}
\item Sub-array -- Defines the size and geometry of the region being read out and the time it takes to read a single frame.
\item Readout pattern -- Defines the number of frames averaged per group and number of frames skipped between groups.
\item \# of groups -- Number of groups per ramp.
\item \# of integrations -- Number of integration ramps.
\item \# of exposures -- Number of individual exposures, i.~e. sets of integrations. 
\end{itemize}

For un-dithered observations, the number of integrations and number of exposures are algorithmically interchangeable.  For dithered observations, the dither specification defines how many dither positions per observation, the number of exposures defines how many observations, and the rest define the time spent at each dither position.  Currently, dithering is only supported for IFU observations.

\subsection{Throughputs and Optical Path}

The throughput conversion uses a standard ETC methodology prior to reaching the detector; each optical element along the path is associated with a scalar or chromatic (one-dimensional) throughput efficiency curve. The individual throughput efficiencies are assumed to be multiplicative. Optical elements common to all instruments include the Optical Telescope Element (OTE), internal mirror reflections, filters, grating efficiencies (for spectroscopic modes), and the quantum efficiency of the detector. Diffractive throughput components, such as pupil stops and the sparse aperture mask of NIRISS, are generally taken into account by the PSF model.

\subsection{Dynamic Range and Saturation}

There are at least two types of saturation for both the near- and mid-infrared detectors: hard and soft. That is, since the readouts are non-destructive, it is possible to measure a ramp slope, even if some reads saturate in the latter part of the ramp. Pandeia calls this {\it soft saturation}. The requirement for a meaningful ramp measurement is that a minimum number of groups are unsaturated. Two groups is currently a strict minimum to define a line, but more may be needed in practice, depending on the number of parameters required for a ramp-fitting algorithm used in calibration. Ramps that saturate in less than the minimum number of reads represent {\it hard saturation}. Therefore, pixels with soft saturation can still be used to construct ETC products (SNRs, etc.) by replacing the number of groups in the ramp noise equation by the effective number of unsaturated groups, $n_{\rm eff}$. If pixels have soft saturation, the ETC will generate a warning describing how many pixels are affected. If the ETC calculation includes pixels with hard saturation, a signal-to-noise ratio cannot be computed for those pixels and a warning is generated to that effect. 

In some cases, it's possible to use a single-read noise model in lieu of fitting a ramp.  This results in a significantly higher level of noise, but could be preferable for rapid cadence observations of very bright sources.  We are planning to implement a single-read option for at least JWST NIRCam and NIRSpec. 

\section{Noise Propagation}

Traditionally, the noise calculation in an exposure time calculator assumes uncorrelated normal distributions for the signal in detector pixels. For the JWST detectors (both at near-infrared and mid-infrared wavelengths), these assumptions do not hold and can lead to estimates of signal-to-noise ratios that are incorrect by significant amounts. Pandeia includes a framework for calculating signal-to-noise ratios for detectors that have correlated pixels. This generalized noise propagation framework also allows for a realistic treatment of different observing strategies and an evaluation of the impact of those strategies on the final data quality.

\subsection{MULTIACCUM Ramp Noise Model and Correlated Noise}

Pandeia adopts the noise formulation of Rauscher (2007)\cite{2007PASP..119..768R} for calculating the noise on a single pixel. The readout noise in the near-infrared HAWAII 2RG detectors used in the three of the four JWST instruments (NIRCam, NIRSpec, and NIRISS) is highly correlated. That is, the signal sampled in one pixel is dependent on what is sampled in other pixels, even in the absence
of incident light onto the detector. The strongest correlation is along the fast-read direction and it generally extends to large distances from any given pixel. The mid-infrared detectors used by MIRI may also be affected by correlated noise, though to a lesser degree. Finally, pixel-to-pixel noise correlation may also be used to describe speckle noise seen in high contrast imaging
applications such as those offered by the various coronagraphic modes.

The immediate consequence of such correlation is that variances do not add in quadrature. That is, the commonly used error propagation of a sum (e.g., of pixels) does not hold:

\begin{equation}
\sigma^2 (x_1+x_2 )\neq \sigma^2 (x_1)+ \sigma^2 (x_2)
\end{equation}

Instead, the noise on the sum is higher with an amount that is the covariance of the two pixels:

\begin{equation}
\sigma^2 (x_1+x_2 ) = \sigma^2 (x_1 )+ \sigma^2 (x_2 )+2R_{12} \sigma^2 (x_1)\sigma^2 (x_2),
\end{equation}

where $R_{12}$ is the correlation coefficient or, more specifically, the Pearson product-moment correlation coefficient. If $R_{12}=0$, there is no correlation. Conversely, if $R_{12}=1$, there is complete correlation, and $\sigma(x_1+x_2)= \sigma_1 (x_1 )+\sigma_2 (x_2 )$.

A general flux measurement on a two-dimensional detector can be expressed as a linear combination of pixel values:

\begin{equation}
F_{tot}=\sum_i{a_i F_i}
\end{equation}

$F_i$ is the flux value in each pixel and $a_i$ is a noiseless scalar weight encapsulating both optimal extraction schemes, as well as background subtraction (in which case $a<0$). The general error propagation law for a linear combination is:

\begin{equation}
\sigma^2 (F_{tot} )=\bar{a}\bar{C}\bar{a}^T,
\label{noise_sum}
\end{equation}

where $\bar{C}$ is the covariance matrix of $\bar{F}$. The covariance matrix is a less general, scaled version of the correlation matrix: $C_{ij} = R_{ij} \sigma_i \sigma_j$. Note that the $(i,j)$ subscripts do not refer to the coordinates of a pixel on a two-dimensional detector. Rather, they are one-dimensional subscripts that can refer to pixel coordinates $(x,y)$, for instance in the following way for a square subarray: $i=x+ny$. If the sum is over an $n\times n$ pixel square, the covariance matrix will have dimensions $n^2\times n^2$.

Thus, for an appropriate calculation of the total noise of the flux of an astronomical source in an image or in a spectral channel, the main difficulty is in constructing an appropriate covariance matrix, which will yield the total variance upon a simple matrix multiplication with a set of pixel-by-pixel weights.

\subsection{The Correlated Read Noise Term}

\begin{figure}[ht!]
\centering
\includegraphics[width=10cm]{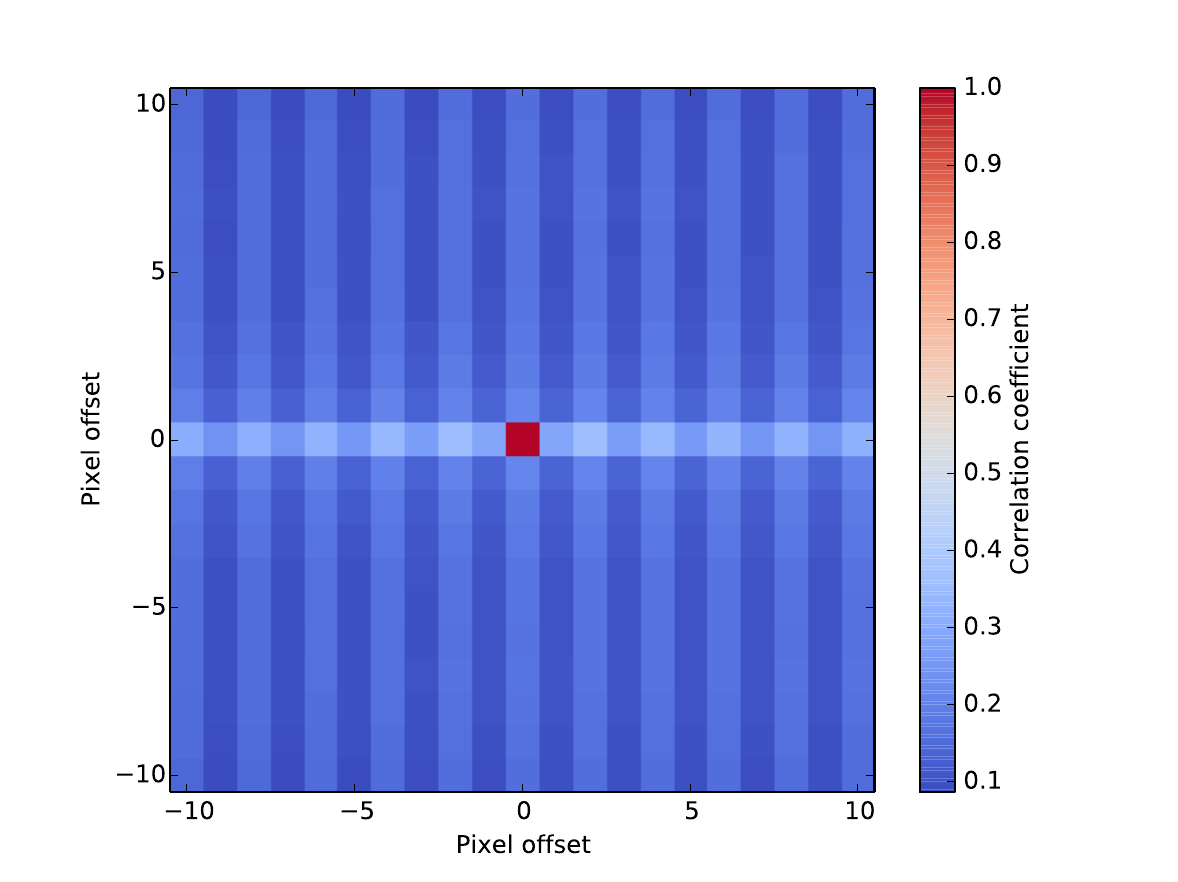}

\caption{Correlation matrix for a HAWAII2RG detector used by the ETC as one input to the covariance matrix. The correlation depends on the number of frames in the integration ramp, and therefore on the readout pattern.}

\label{CorrMatrix}
\end{figure}

The single-pixel read noise and correlation between pixels is assumed to be independent of the pixel. That is: $C_{ij}=R_{ij} \sigma_{RN}^2$. The correlation matrix $R_{ij}$ can be constructed from the correlation of a single pixel in detector coordinates $R_{\Delta x\Delta y}$: $R_{ij}=T(R_{\Delta x\Delta y})$. See Figure \ref{CorrMatrix} for an example of a notional HAWAII 2RG read noise correlation matrix. Calculating the read noise for a given extraction strategy an aperture is then accomplished by setting the weights, $\bar{a}$, appropriately, that is for pixels outside the aperture, $a_i=0$, while for background pixels intended to be subtracted, $a<0$. In summary, we can define a constant covariance matrix, $C_{RN}$, such that:

\begin{equation}
\sigma_{RN}^2=\bar{a}\bar{C_{\rm RN}} \bar{a}^T
\end{equation}

The covariance matrix can be stored as a reference file, relevant for one detector type, and possibly for each readout and sub-array configuration.

\subsection{Inter-Pixel Capacitance (IPC)}

The other source of correlated noise is related to a correlation of the signal in one pixel to those of adjacent pixels. For the JWST near-infrared detectors, this correlation is carried primarily by the inter-pixel capacitance (IPC) effect\cite{2004SPIE.5167..204M}. The IPC leads to the detection of charge that accumulated in adjacent pixels along columns and rows (forming a characteristic ``$+$'', formed by a set of 5 pixels). This is not a stochastic effect (such as charge diffusion), but is completely deterministic. That means that the Poisson noise is correlated between adjacent pixels. As an example, consider a dark array in which only a single pixel is exposed to light. A fraction $f$ of the signal $S$ (in electrons) in that pixel is
detected in the 4 adjacent pixels (the diagonal pixels are assumed to be uncoupled). If $f=0$, the Poisson noise of the central pixel is the classical relation:

\begin{equation}
\sigma_{\rm Poisson}^2 = S
\end{equation}

However, if $f>0$, this relation still holds for the central pixel, that is, even though some of the signal was detected in adjacent pixels, the Poisson noise is still that which the pixel should have had in the absence of IPC. Thus, it is an underestimate to take $\sigma_{\rm Poisson}^2=S(1-f)$. This leads to problems when using differences of flat fields to determine
the detector gain and other fundamental parameters. However, in an astronomical application, which would sum over the 5 pixels with signal, the Poisson noise of the sum remains $\sigma_{\rm Poisson}^2=S= \Sigma s_i$. Errors in the Poisson noise estimate occurs if some of the signal is not included in the sum, for instance in aperture photometry.

In an actual astronomical observation, the correction for IPC effects on the Poisson noise estimate thus essentially requires a deconvolution of the image. However, in an ETC model, we already have the IPC deconvolved image. The noise of a strategy product is therefore calculated using an unconvolved detector image, while the signal calculation is performed on a detector image that has been convolved with the IPC kernel.

\subsection{Cosmic Rays}
Cosmic rays are a significant source of noise for any space-based observatory, including those located at L2\cite{2014SPIE.9154E..06K}\cite{2012arXiv1207.5597H} where JWST and WFIRST will be stationed. The ability to fit ramps to MULTIACCUM data provides some robustness against the effects of cosmic rays. A slope can still be determined for the portion of a ramp before the
cosmic ray and sometimes afterward as well. 

Pandeia treats cosmic rays effects on an average basis and assumes conservatively that only the portion of the ramp before a cosmic ray event can still be used.  Under this assumption, the average integration time for a CR-affected ramp is:

\begin{equation}
t_{ave} = \frac{(n_{read} - n_{min})}{2(n_{read}-1)} t_{ramp}
\end{equation}

where $t_{ramp}$ is the total integration time for the ramp, $n_{read}$ is the number of readouts along the ramp, and $n_{min}$ is the minimum number of reads required to determine a valid slope. From this, an effective mean exposure time for a ramp subject to a per-pixel cosmic ray rate of $R_{cr}$ is [update note: the SPIE manuscript contains an error in the formula below. This version contains the correction.]:

\begin{equation}
t_{eff} = t_{ramp}\Big[ 1.0 - \Big( 1.0 - \frac{(n_{read} - n_{min})}{2(n_{read}-1)}\Big) R_{cr} t_{ramp}\Big]
\end{equation}

\section{Extraction Strategies}

The error propagation formalism described above is semantically the implementation of a so-called observing {\it strategy} on a set of pixel flux measurements (typically arranged as a two-dimensional image). The strategy describes how a map of detector pixel electron rates and noise rates is transformed unto a scientific product, usually, but not limited to, a signal, noise,  and signal-to-noise ratio of a photometric point or extracted spectrum. Given an observing strategy, the software implementation of the process can be condensed to the construction of 1) the read noise covariance matrix, given a set of read-out parameters and 2) a weight map $a_i$, where the index $i$ runs over all the pixels involved in an extraction from a 2D detector rate image to a 1D spectrum or a scalar photometric point.

\subsection{Imaging Aperture Photometry}

As an illustrative example, one of the simplest observing strategies is that of imaging aperture photometry. In this case, we perform a simple sum of pixels inside some aperture, $A$. Further, from each of the aperture pixels we subtract a background value, which is estimated as a mean of a set of pixels within a background aperture, $B$:

\begin{equation}
F=K \sum_i^{N_A}{ (F_i^{\rm aperture}-1/N_B  \sum_j^{N_B}{F_j^{\rm background}} ) }
\label{apphot}
\end{equation}

$K$ is a constant that corrects the total flux in the case that the aperture does not include all of the source flux. In a signal-to-noise ratio, this constant will drop out. Clearly, this formula satisfies the condition that our measurement is constructed as a linear combination of pixel values, with a weight map, provided $A\cap B=\emptyset$:

\begin{equation}
a_i = \begin{cases}
      K,         & i \in A \\
      -KN_A/N_B, & i \in B
      \end{cases}
\end{equation}

Inputting the resulting $a_i$ vector into Eq.~\ref{noise_sum} yields the variance of the photometry and is returned by the ETC as a signal-to-noise ratio.

\subsection{Spectroscopic Aperture Photometry}

For dispersed modes, the aperture photometry is performed at each wavelength step along the dispersion axis.  In this case, Eq.~\ref{apphot} yields an extracted spectrum in the form of an array of flux values and applying Eq.~\ref{noise_sum} yields an array of signal-to-noise ratios.  Specific dispersed modes vary largely in how their source and background apertures are defined.

\subsubsection{Fixed Slit}
The simplest spectroscopic case extracts flux at each wavelength step from a 1D extraction aperture and, optionally, subtracts background estimated from one or more non-overlapping 1D background apertures. The NIRSpec MSA strategy is based on this, but with the sizes and locations of the source and background apertures constrained by the size and spacing of MSA apertures.

\subsubsection{Slitless}
In slitless cases, including multi-order slitless, the flux is extracted at each wavelength step via 1D apertures, but the background is estimated from 2D regions. Slitless background is not wavelength dependent so larger, 2D estimation regions can be used to reduce the noise in the measurement of the background flux and shared for all wavelengths.

\subsubsection{Integral Field}
For integral field spectroscopy, the extraction apertures are defined as 2D shapes in spatial coordinates which are then mapped into detector pixels according to the IFU slice configuration. Flux is extracted at each wavelength step and there are three methods for estimating the background:
\begin{itemize}
\item {\em Aperture} -- Analogous to imaging aperture photometry, define one or more 2D regions and estimate the background from them for each wavelength step.
\item {\em Dither Off-Scene} -- Interleave on-source and off-source exposures and subtract normalized average of off-source exposures from on-source ones. This has the effect of more than doubling the time required to perform an observation, but will likely be used often in practice to subtract background from sources that significantly fill an IFU field-of-view. 
\item {\em Dither On-Scene} -- Interleave exposures centered at two different positions within a scene and use the same extraction aperture definition to extract signal and background from each exposure. This method is appropriate for sources that are much smaller than the IFU field-of-view and realistically models the effects of background contamination by other sources in the scene.
\end{itemize}

\section{Performance and benchmarks}
The JWST ETC has been extensively tested in preparation for release. The testing has been separated into two steps. First, the algorithms and software implementation were benchmarked relative to sensitivity calculations provided by the JWST instrument teams, assuming the same or similar reference data (throughputs, background, etc.). Since there are aspects of the pixel-based approach that cannot be directly compared to one-dimensional sensitivity estimates, the benchmarks comparisons were generally required to match to $\sim 10\%$ in derived signal-to-noise ratios. The second step is to ensure that the ETC is using the best known reference data. This process is currently ongoing. 

\section{Pandeia for WFIRST and Other Missions}

Pandeia is fully data-driven and a new observatory and instrument suite can be implemented largely by adding a new set of configuration files and appropriate reference data. All of the JWST instruments and observing modes are currently implemented in preparation for the first JWST call for proposals. Support has also been implemented for the current designs for the WFIRST Wide-field Imager (WFI) and integral field channels\cite{WFIRST}. Support for the WFIRST coronagraph is planned once a baseline design for the instrument is finalized.

\section{Accessing and Using Pandeia}
\begin{figure}[ht!]
\centering
\includegraphics[width=15cm]{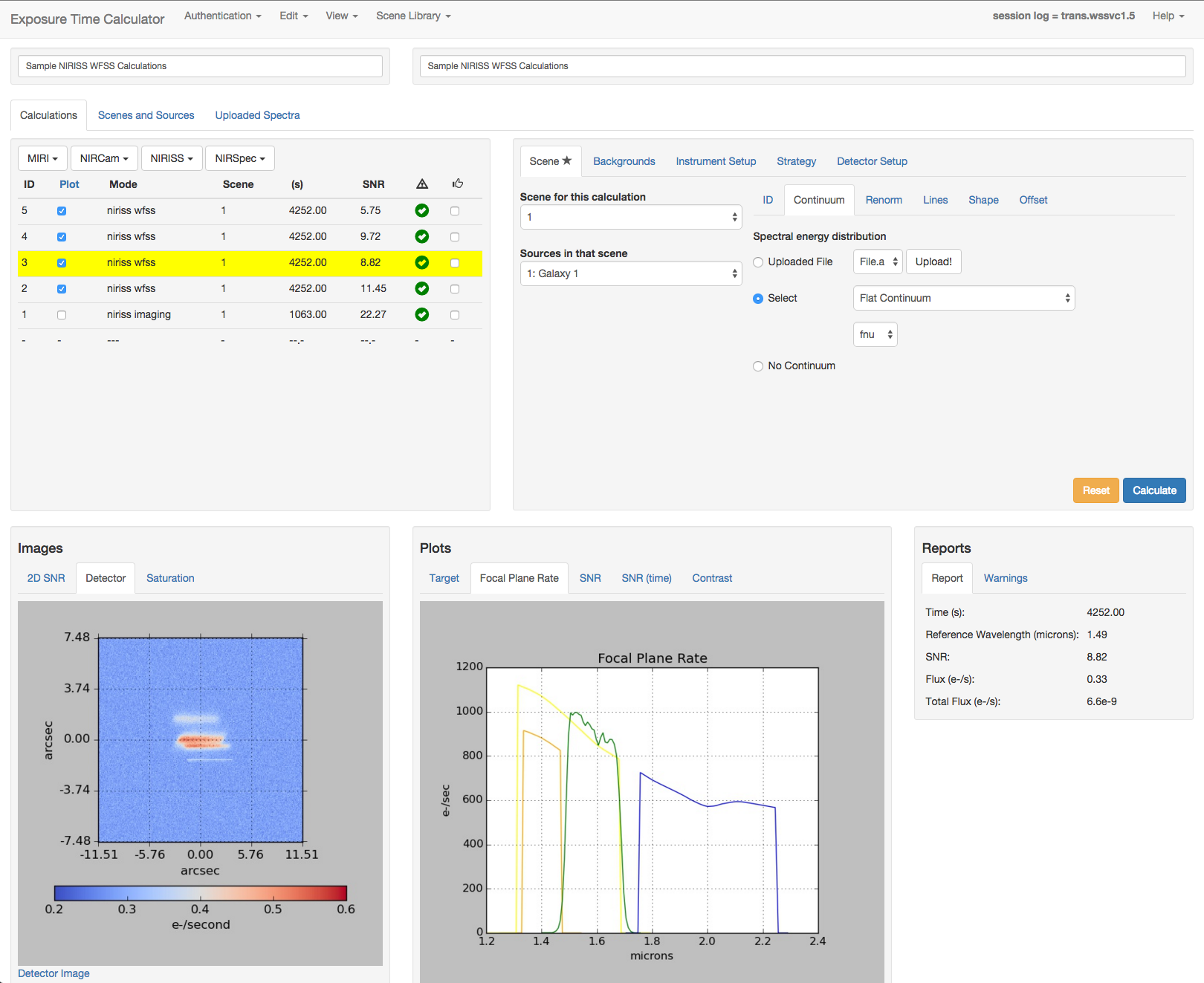}
\caption{Representative screenshot of the JWST web-interface to the Pandeia ETC.}
\label{etcpage}
\end{figure}

For JWST users, the primary method of using Pandeia will be via a modern web-based interface that is currently under development.  It will provide much more sophisticated means for grouping calculations into ``workbooks'', sharing configuration data for objects or groups of objects between calculations, and sharing workbooks with other users. It will also provide richer visualization of 2D and 3D ETC outputs as well as the ability to superimpose results from multiple calculations for easy comparison. A representative screenshot is shown in Figure~\ref{etcpage}. 

An initial limited-support, pre-release of the JWST web interface is expected to made available to the public by late summer 2016. The official public release to support preparations for Cycle 1 is scheduled for January 2017. The underlying, Python-based compute engine will also be made public via STScI's AstroConda channel\cite{astroconda} as of summer 2016.  Up-to-date information and links to access the JWST ETC are available at \url{https://jwst.stsci.edu/science planning/performance--simulation-tools-1/exposure-time-calculator-etc}. 

For WFIRST users, a limited WFIRST-only version of the compute engine has been available since January 2016\cite{WFIRST}. It has been only available to members of the WFIRST team per-request with all computations performed via a JupyterHub\cite{jupyter} notebook server hosted at STScI.  While not as user-friendly as the web interface being developed, Jupyter notebooks provide a powerful way for experienced users to explore ETC capabilities in a way that can be reproduced and shared with others. We will include some example notebooks with the compute engine when it is publicly released. 

\acknowledgments 
This work made use of Astropy, a community-developed core Python package for Astronomy (Astropy Collaboration, 2013\cite{2013AA...558A..33A}).

\bibliography{report} 

\begin{thebibliography}{10}

\bibitem{1999ApJ...512..377H}
{Hauschildt}, P.~H., {Allard}, F., and {Baron}, E., ``{The NextGen Model
  Atmosphere Grid for Teff = 3000--10,000 K},'' {\em ApJ}~{\bf 512},  377--385
  (Feb. 1999).

\bibitem{0067-0049-212-2-18}
Brown, M. J.~I., Moustakas, J., Smith, J.-D.~T., da~Cunha, E., Jarrett, T.~H.,
  Imanishi, M., Armus, L., Brandl, B.~R., and Peek, J. E.~G., ``An atlas of
  galaxy spectral energy distributions from the ultraviolet to the
  mid-infrared,'' {\em APJS}~{\bf 212}(2),  18 (2014).

\bibitem{2001ApJ...548..296W}
{Weingartner}, J.~C. and {Draine}, B.~T., ``{Dust Grain-Size Distributions and
  Extinction in the Milky Way, Large Magellanic Cloud, and Small Magellanic
  Cloud},'' {\em ApJ}~{\bf 548},  296--309 (Feb. 2001).

\bibitem{2007ApJ...663.1069F}
{Flaherty}, K.~M., {Pipher}, J.~L., {Megeath}, S.~T., {Winston}, E.~M.,
  {Gutermuth}, R.~A., {Muzerolle}, J., {Allen}, L.~E., and {Fazio}, G.~G.,
  ``{Infrared Extinction toward Nearby Star-forming Regions},'' {\em ApJ}~{\bf
  663},  1069--1082 (July 2007).

\bibitem{2009ApJ...691..560C}
{Chapman}, S.~C., {Blain}, A., {Ibata}, R., {Ivison}, R.~J., {Smail}, I., and
  {Morrison}, G., ``{Do Submillimeter Galaxies Really Trace the Most Massive
  Dark-Matter Halos? Discovery of a High-z Cluster in a Highly Active Phase of
  Evolution},'' {\em ApJ}~{\bf 691},  560--568 (Jan. 2009).

\bibitem{2012SPIE.8442E..3DP}
{Perrin}, M.~D., {Soummer}, R., {Elliott}, E.~M., {Lallo}, M.~D., and
  {Sivaramakrishnan}, A., ``{Simulating point spread functions for the James
  Webb Space Telescope with WebbPSF},'' in [{\em Space Telescopes and
  Instrumentation 2012: Optical, Infrared, and Millimeter
  Wave}{\nolinebreak\hspace{0.1em}]},  {\em Proc. SPIE} {\bf 8442},  84423D
  (Sept. 2012).

\bibitem{2014SPIE.9143E..3XP}
{Perrin}, M.~D., {Sivaramakrishnan}, A., {Lajoie}, C.-P., {Elliott}, E.,
  {Pueyo}, L., {Ravindranath}, S., and {Albert}, L., ``{Updated point spread
  function simulations for JWST with WebbPSF},'' in [{\em Space Telescopes and
  Instrumentation 2014: Optical, Infrared, and Millimeter
  Wave}{\nolinebreak\hspace{0.1em}]},  {\em Proc. SPIE} {\bf 9143},  91433X
  (Aug. 2014).

\bibitem{2012SPIE.8442E..2RD}
{Doyon}, R., {Hutchings}, J.~B., {Beaulieu}, M., {Albert}, L.,
  {Lafreni{\`e}re}, D., {Willott}, C., {Touahri}, D., {Rowlands}, N.,
  {Maszkiewicz}, M., {Fullerton}, A.~W., {Volk}, K., {Martel}, A.~R., {Chayer},
  P., {Sivaramakrishnan}, A., {Abraham}, R., {Ferrarese}, L., {Jayawardhana},
  R., {Johnstone}, D., {Meyer}, M., {Pipher}, J.~L., and {Sawicki}, M., ``{The
  JWST Fine Guidance Sensor (FGS) and Near-Infrared Imager and Slitless
  Spectrograph (NIRISS)},'' in [{\em Space Telescopes and Instrumentation 2012:
  Optical, Infrared, and Millimeter Wave}{\nolinebreak\hspace{0.1em}]},  {\em
  Proc. SPIE} {\bf 8442},  84422R (Sept. 2012).

\bibitem{2007PASP..119..768R}
{Rauscher}, B.~J., {Fox}, O., {Ferruit}, P., {Hill}, R.~J., {Waczynski}, A.,
  {Wen}, Y., {Xia-Serafino}, W., {Mott}, B., {Alexander}, D., {Brambora},
  C.~K., {Derro}, R., {Engler}, C., {Garrison}, M.~B., {Johnson}, T.,
  {Manthripragada}, S.~S., {Marsh}, J.~M., {Marshall}, C., {Martineau}, R.~J.,
  {Shakoorzadeh}, K.~B., {Wilson}, D., {Roher}, W.~D., {Smith}, M., {Cabelli},
  C., {Garnett}, J., {Loose}, M., {Wong-Anglin}, S., {Zandian}, M., {Cheng},
  E., {Ellis}, T., {Howe}, B., {Jurado}, M., {Lee}, G., {Nieznanski}, J.,
  {Wallis}, P., {York}, J., {Regan}, M.~W., {Hall}, D.~N.~B., {Hodapp}, K.~W.,
  {B{\"o}ker}, T., {De Marchi}, G., {Jakobsen}, P., and {Strada}, P.,
  ``{Detectors for the James Webb Space Telescope Near-Infrared Spectrograph.
  I. Readout Mode, Noise Model, and Calibration Considerations},'' {\em
  PASP}~{\bf 119},  768--786 (July 2007).

\bibitem{2004SPIE.5167..204M}
{Moore}, A.~C., {Ninkov}, Z., and {Forrest}, W.~J., ``{Interpixel capacitance
  in nondestructive focal plane arrays},'' in [{\em Focal Plane Arrays for
  Space Telescopes}{\nolinebreak\hspace{0.1em}]},  {Grycewicz}, T.~J. and
  {McCreight}, C.~R., eds., {\em Proc. SPIE} {\bf 5167},  204--215 (Jan. 2004).

\bibitem{2014SPIE.9154E..06K}
{Kohley}, R., {Crowley}, C.~M., {Gar{\'e}}, P., {Chassat}, F., {Short}, A.~D.,
  {Martin-Fleitas}, J.~M., {Mora}, A., {Abreu-Aramburu}, A., and {Prod'homme},
  T., ``{The radiation environment at L2 as seen by Gaia},'' in [{\em High
  Energy, Optical, and Infrared Detectors for Astronomy
  VI}{\nolinebreak\hspace{0.1em}]},  {\em Proc. SPIE} {\bf 9154},  915406 (July
  2014).

\bibitem{2012arXiv1207.5597H}
{Horeau}, B., {Boulade}, O., {Claret}, A., {Feuchtgruber}, H., {Okumura}, K.,
  {Panuzzo}, P., {Papageorgiou}, A., {Rev{\'e}ret}, V., {Rodriguez}, L., and
  {Sauvage}, M., ``{Impacts of The Radiation Environment At L2 On Bolometers
  Onboard The Herschel Space Observatory},'' {\em ArXiv e-prints}  (July 2012).

\bibitem{WFIRST}
``Wide field infrared survey telescope source simulation and exposure time
  calculation.'' \url{http://www.stsci.edu/wfirst/software/Pandeia} (2016).

\bibitem{astroconda}
``Astroconda.'' \url{http://astroconda.readthedocs.io/en/latest/} (2016).

\bibitem{jupyter}
``Jupyterhub.'' \url{http://jupyterhub.readthedocs.io/en/latest/} (2016).

\bibitem{2013AA...558A..33A}
{Astropy Collaboration}, {Robitaille}, T.~P., {Tollerud}, E.~J., {Greenfield},
  P., {Droettboom}, M., {Bray}, E., {Aldcroft}, T., {Davis}, M., {Ginsburg},
  A., {Price-Whelan}, A.~M., {Kerzendorf}, W.~E., {Conley}, A., {Crighton}, N.,
  {Barbary}, K., {Muna}, D., {Ferguson}, H., {Grollier}, F., {Parikh}, M.~M.,
  {Nair}, P.~H., {Unther}, H.~M., {Deil}, C., {Woillez}, J., {Conseil}, S.,
  {Kramer}, R., {Turner}, J.~E.~H., {Singer}, L., {Fox}, R., {Weaver}, B.~A.,
  {Zabalza}, V., {Edwards}, Z.~I., {Azalee Bostroem}, K., {Burke}, D.~J.,
  {Casey}, A.~R., {Crawford}, S.~M., {Dencheva}, N., {Ely}, J., {Jenness}, T.,
  {Labrie}, K., {Lim}, P.~L., {Pierfederici}, F., {Pontzen}, A., {Ptak}, A.,
  {Refsdal}, B., {Servillat}, M., and {Streicher}, O., ``{Astropy: A community
  Python package for astronomy},'' {\em A\&A}~{\bf 558},  A33 (Oct. 2013).

\end{thebibliography}
\bibliographystyle{spiebib} 

\end{document}